\documentclass[a4paper]{jpconf}
\usepackage{graphicx,hyperref, amsmath, amssymb}
\begin{document}
\title{Fluctuations of Strongly-interacting Matter in Thermal Models at Chemical Freeze-out}

\author{Jamie M. Karthein}

\address{Center for Theoretical Physics, Massachusetts Institute of Technology, Cambridge, MA, 02139, USA}

\ead{jmkar@mit.edu}

\begin{abstract}
Fluctuations provide a powerful tool for elucidating the nature of strongly-interacting matter in the QCD phase diagram. In heavy-ion-collision systems, the net-particle number fluctuations are captured at the moment of chemical freeze-out. Studies of the chemical freeze-out via susceptibilities from lattice QCD and the Hadron Resonance Gas model contribute to the characterization of the transition region of the QCD phase diagram. This contribution to proceedings will show how susceptibilities can be used to study the interplay between different conserved charges via cross-correlators and to constrain interactions in the hadron gas phase.
\end{abstract}

\section{Introduction} \label{sec:intro}
In the characterization of QCD thermodynamics, the measured fluctuations of and correlations between particles offer an important connection between experimental data and theoretical calculations.
From the theory perspective, the fluctuations correspond to the susceptibilities, which are defined as derivatives of the pressure with respect to the conserved charge chemical potentials, as shown in Eq. \eqref{eq:susc}.
\begin{equation}
    \chi_{ijk}^{BQS} = \frac{\partial^{i+j+k} (P/T^4)}{\partial(\mu_B/T)^i \partial(\mu_Q/T)^j \partial(\mu_S/T)^k}
    \label{eq:susc}
\end{equation}
Many calculations of the fluctuations have been performed over the year with first-principles lattice QCD techniques, including diagonal \cite{Borsanyi:2011sw,Borsanyi:2014ewa,Bellwied:2015lba,Noronha-Hostler:2016rpd,Bazavov:2017tot} and off-diagonal \cite{Borsanyi:2018grb,Bazavov:2020bjn} fluctuations of conserved charges.

On the experimental side, the fluctuations are manifested in the statistical distribution of the net number of particles, i.e. the difference of particles minus antiparticles, carrying the conserved charge of choice. 
Given the fact that there are millions of events being measured in heavy-ion-collision (HIC) experiments, the natural way to characterize the change in the results on an event-by-event basis is through the fluctuations, or moments, of the distribution: mean, variance, skewness, and kurtosis. 
The relationship between the experimental moments and the susceptibilities is given by Eq. \eqref{eq:moments}.
\begin{align}
\mathrm{ mean:}&~~M=\langle N\rangle= VT^3\chi_1\nonumber\,,\\
\mathrm{ variance:}&~~\sigma^2=
\langle (\delta N)^2\rangle=VT^3\chi_2
\nonumber\,,
\\
\mathrm{ skewness:}&~~S=\frac{\langle (\delta N)^3\rangle}{\sigma^3}=\frac{VT^3\chi_3}{(VT^3\chi_{2})^{3/2}}
\nonumber\,,
\\
\mathrm{kurtosis:}&~~\kappa=\frac{\langle (\delta N)^4\rangle}{\sigma^4}-3=\frac{VT^3\chi_4}{(VT^3\chi_{2})^{2}}\,\,,
\label{eq:moments}
\end{align}
where $\delta N=N-\langle N\rangle$.
In order to eliminate the volume dependence to leading order, ratios of the moments can be constructed, such that the susceptibilities from the theory can then be compared to the experimental moments.
\begin{align}
\sigma^2/M&=\chi_2/\chi_1\,,~~~~~~~~~~~~S\sigma=\chi_3/\chi_{2}\,,
\nonumber\\
\kappa\sigma^2&=\chi_4/\chi_{2}\,,~~~~~~S\sigma^3/M=\chi_3/\chi_1\,.
\label{eq:moment-susc-ratios}
\end{align}

Similarly, the correlations, which describe the interactions between two or more particles, are also defined in terms of the susceptibilities. 
In fact, it is possible to construct a matrix that allows for the identification of second-order diagonal and off-diagonal (or cross) correlators of $BQS$ conserved charges:
\begin{align*}
\begin{pmatrix}
    \chi^2_B & \chi^{11}_{BQ} & \chi^{11}_{BS} \\
    \chi^{11}_{BQ} & \chi^2_Q & \chi^{11}_{QS} \\
    \chi^{11}_{BS} & \chi^{11}_{QS} & \chi^2_S\\
\end{pmatrix},
\end{align*}
where the cross-correlators are given by mixed derivatives:
\begin{equation}
    \chi^{11}_{ij}= \frac{\partial^2(P/T^4)}{\partial(\mu_i/T)\partial(\mu_j/T)} \, .
    \nonumber
\end{equation}

The main focus of this contribution is to use fluctuations to explore the chemical freeze-out stage in heavy-ion-collisions that occurs after the phase transition from deconfined quark matter to hadronic degrees of freedom.
While the fluctuations measured by the experiment are given for net-particle number, the ones from fundamental lattice QCD calculations are in terms of the conserved charges for strong interactions. 
This difference arises from the different systematics for theory and experiment.
On the one hand, the first-principles calculations rely on simulations of discretized configurations of the field theory for which the degrees of freedom are quarks and gluons and which must obey strong interaction conservation of $BQS$.
As such, the theoretical system does not have access to individual particle species since the hadron yields are not a conserved charge.
On the other hand, the detectors in the experiments are measuring particles, and in particular, only a subset of all hadronic species created during the collisions due to the limitations of the detectors, e.g. the insensitivity to electrically neutral species like the neutron.
In order to bridge the differences in the treatment of the system between theory and experiment, the use of a robust, adaptable model is necessary.

A powerful framework for understanding the matter created in heavy-ion collisions is the Hadron Resonance Gas (HRG) model.
The pressure in the HRG model is calculated as a sum over the species in a given hadronic list. 
\begin{equation}
\label{eq:HRG_press}
\centering
    P(T,\mu_B,\mu_Q,\mu_S) = \sum_{i \in \text{HRG}} \pm \frac{d_i T}{(2\pi)^3} \int d^3 p \ln  \small( 1 \pm \text{exp} \small[ - \small(\sqrt{\Vec{p^2} + m^2}  - B_i \mu_B - Q_i \mu_Q - S_i \mu_S \small)/T \small] \small)
\end{equation}
%
Due to the difference in particle nature of the mesons and baryons, they must each be treated with either Bose-Einstein or Fermi-Dirac statistics, respectively, for which the former appears with a ($-$) sign, while the latter carries a (+) sign.
The main input for the model is the list of particles that have individual degeneracy $d_i$, mass $m_i$, and quantum numbers $B_i$, $S_i$, and $Q_i$.
As we can see in Eq. \ref{eq:HRG_press}, any additional states seek to increase the pressure of the system, i.e. a larger number of states will lead to a larger overall pressure.
The chemical potentials $\mu_B$, $\mu_Q$, and $\mu_S$ are all linked due to the experimental conditions for heavy-ion collisions.
Namely, they exhibit global net-strangeness neutrality and an approximate ratio of 40\% protons to total nucleons (baryons) in the colliding nuclei such that:
\begin{align}
\label{eq:str_neut}
\centering
    \langle n_S (T, \,\mu_B, \,\mu_Q, \,\mu_S) \rangle &= 0 \nonumber
    \\
    \langle n_Q (T, \,\mu_B, \,\mu_Q, \,\mu_S) \rangle &= \frac{Z}{A} \langle n_B(T, \,\mu_B, \,\mu_Q, \,\mu_S) \rangle,
\end{align}
where $Z/A$ in heavy-ion collisions is $\sim 0.4$, e.g. at the LHC $(Z/A)_\text{Pb} = 82/208$.

\section{Correlators of Conserved Charges in QCD}

Particle correlators are important for the study of chemical freeze-out in heavy-ion collisions.
These correlations represent the interactions between two or more particles, including self-interactions and interactions amongst different types of particles.
Correlations can be understood as the mechanism by which we can quantify to what extent particles ``feel" one another.
Correlators of conserved charges can show the intrinsic (self-)correlations that are present in QCD by considering (diagonal) off-diagonal susceptibilities.
Furthermore, correlations of conserved charges provide an insight into the interplay between the different $BQS$ charges.
These quantities are truly sensitive to the intrinsic nature of strongly-interacting matter due to the fact that quarks carry more than one conserved charge.
For example, the strange quark has fractional quantum numbers $\frac{1}{3}B$, $-\frac{1}{3}Q$, and $-S$, while the up quark carries $\frac{1}{3}B$ and $\frac{2}{3}Q$.
Originally, the interest in conserved charge correlators came from the study of a baryon-strangeness correlator \cite{Koch:2005vg}.
This study proposed the construction of the ratio $C_{BS}=-3\chi^{BS}_{11}/\chi^S_2$ as a baryon-strangeness correlator which was shown to be sensitive to the transition to deconfined matter.
However, it is difficult to confirm the quantity experimentally since the hadron yields measured by the experiment do not capture all the species contributing to the baryon and strangeness conserved charges.
It was, thus, necessary to employ the HRG model in order to determine how much of a given correlator can be seen in the experimental data, as in the work published in Ref. \cite{Bellwied:2019pxh}.

The framework for the HRG model employed here is described by Eq. \eqref{eq:HRG_press}.
The conserved charge correlators follow from the definition of the susceptibilities given in Eq. \eqref{eq:susc}.
The goal of this particular study is to construct correlators that are simultaneously experimentally accessible and carry the majority of the signal for the conserved charges.
In order to do so, only species stable to strong hadronic decays are considered within the HRG model. 
On the other hand, experimental measurements are limited, essentially to charged particles, and so we cannot access all hadrons relevant for every conserved charge. 
Typically, approximate quantities, so-called proxies, for conserved charges are considered by the experiment in the following way: protons for baryon number, kaons for strangeness, and the sum of measured $\pi$,  $K$, and $p$ for electric charge.
In this analysis, the species  we consider as experimentally measured are:
\begin{equation*}
    \pi^\pm, K^\pm, p, \bar{p}, \Lambda, \bar{\Lambda}, \Xi^-, \Xi^+, \Omega^-, \Omega^+
\end{equation*}
These are of particular interest because, although some others are potentially measurable, results for their yields or fluctuations are not routinely performed at both RHIC and the LHC. 
The inability to detect certain particles leads to a loss of conserved charge in those unmeasured particle species. 

In our framework of the HRG model, we write the pressure derivatives in terms of only the stable hadronic states, as shown in Eq. \eqref{eq:HRG_stable}.  
The sum over all particles is now a double sum over all particles and those stable under strong interactions,
\begin{equation}
    \centering
    \sum_R B_R^i Q_R^j S_R^k \frac{\partial^l(P_R/T^4)}{\partial(\mu_R/T)^l} \rightarrow \sum_{m \in \text{stable}} \sum_R \langle N_m \rangle_R ~B_m^i Q_m^j S_m^k \frac{\partial^l(P_m/T^4)}{\partial(\mu_m/T)^l}
    \label{eq:HRG_stable}
\end{equation}
with $i+j+k=l$, and where the first sum on the right hand side runs over the stable hadrons, and the sum $\langle N_m \rangle_R = \sum_c b^R_c N^R_{m,c}$ gives the average number of particles $m$ produced by each decaying particle R after the entire decay chain. 
Here, the sum runs over particle $R$ decay modes, where $b^R_c$ is the branching ratio of the mode $c$, and $N^R_{m,c}$ is the number of particles $m$ produced by a particle $R$ in the channel $c$.

Next, we write the hadronic contributions to the various conserved charges. 
Furthermore, we will construct net particle ($\tilde{A}$) quantities as $\tilde{A} = A - \bar{A}$ .
With this we define the net BQS conserved charges in the ideal HRG model as follows:
\begin{align}
    \centering
    \text{net-}B &= \tilde{p} + \tilde{n} + \tilde{\Lambda} + \tilde{\Sigma}^+ - \tilde{\Sigma}^- + \tilde{\Xi}^0 + \tilde{\Xi}^- + \tilde{\Omega}^- \nonumber
    \\
    \text{net-}Q &=  \tilde{\pi}^+ + \tilde{K}^+ + \tilde{p} + \tilde{\Sigma}^+ - \tilde{\Sigma}^- - \tilde{\Xi}^- - \tilde{\Omega}^- \nonumber
    \\
    \text{net-}S &=  \tilde{K}^+ + \tilde{K}^0 - \tilde{\Lambda} - \tilde{\Sigma}^+ - \tilde{\Sigma}^- + 2\tilde{\Xi}^0 - 2\tilde{\Xi}^- - 3\tilde{\Omega}^- \, \, .
    \label{eq:conserved_ch}
\end{align}
Given this scheme, the correlators utilize these net-conserved charge quantities, e.g. the $BQ$ correlator:
\begin{equation}
    \centering
    \chi_{11}^{BQ}(T,\mu_B,\mu_Q,\mu_S) = \sum_R (X_{R \rightarrow \text{net-}B}) (X_{R \rightarrow \text{net-}Q}) \frac{\partial^2(P_R/T^4)}{\partial(\mu_B/T)\partial(\mu_Q/T)},
    \label{eq:BQ_corr}
\end{equation}
where $X_{R \rightarrow \text{net-}B} = X_{R \rightarrow \tilde{p}} + X_{R \rightarrow \tilde{n}} + X_{R \rightarrow \tilde{\Lambda}} + X_{R \rightarrow \tilde{\Sigma}^+} + X_{R \rightarrow \tilde{\Sigma}^-} + X_{R \rightarrow \tilde{\Xi}^0} + X_{R \rightarrow \tilde{\Xi}^-} + X_{R \rightarrow \tilde{\Omega}^-}$ and furthermore, $X_{R \rightarrow \tilde{p}} = X_{R \rightarrow p} - X_{R \rightarrow \bar{p}}$. Similar expressions also exist for the various other combinations of $BQS$ conserved charges.

\begin{figure}
	\centering
	\includegraphics[width=\textwidth]{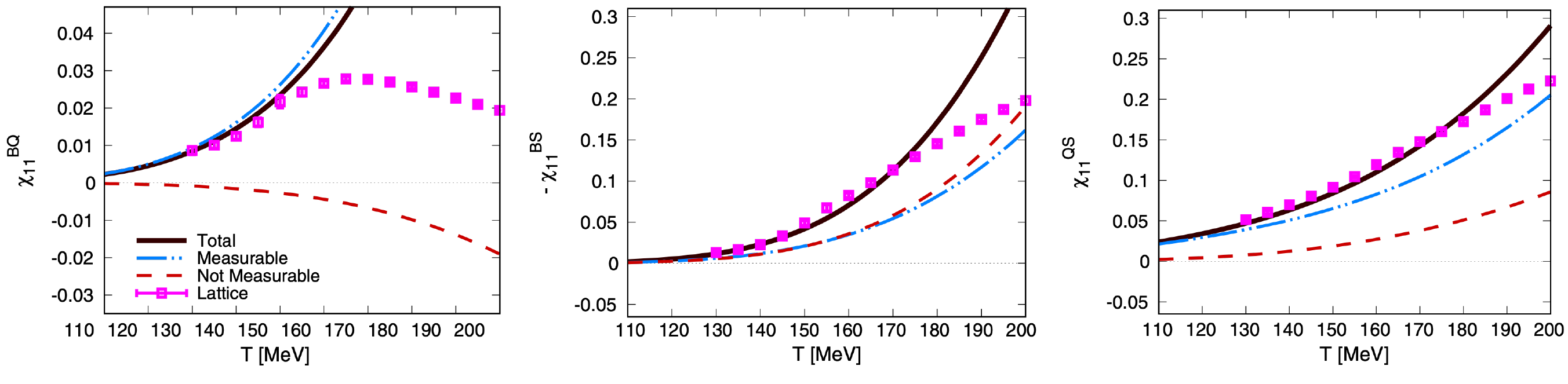}
	\caption[Off-diagonal correlators calculated on the lattice in the continuum limit and compared to HRG model calculations.]{Off-diagonal correlators calculated on the lattice in the continuum limit and compared to HRG model calculations. The black solid curve contains all particles that carry the conserved charges probed by the correlator and is shown to agree with the calculations on the lattice in the continuum limit shown as pink points with error bars. Additionally, the separation into states that the experiment can (dash-dotted curve) and cannot measure (dashed curve) are also calculated within the HRG model \cite{Bellwied:2019pxh}.}
	\label{fig:correl-w-lattice}
\end{figure}

Each of the correlators of conserved charges receives contributions from many different correlations between particle species.  
Furthermore, the sum of these individual particle correlations makes up the measured part of a specific correlator, given that the species are considered ``measured," as defined here. 
In Fig. \ref{fig:correl-w-lattice}, the off-diagonal, or cross, correlators are shown as a function of the temperature at vanishing chemical potential. 
The full result for these correlators, along with the measured and non-measured contributions, from the HRG model are shown in comparison to the
continuum extrapolated data from lattice QCD.


From this plot, we see that the $BQ$ and $QS$ correlators are well-captured by the measured contribution,  while the $BS$ correlator is split between measured and non-measured terms.  
In fact, the measurable part of the $BQ$ correlator even exceeds the full one because, in this case, the non-measured contribution is negative. 
Furthermore, the fact that $BQ$ and $QS$ are well-described by the measured contributions is due to the fact that they are dominated by terms from net-protons and net-kaons, respectively, which in this temperature range form the bulk of particle production, along with the pions. 
On the other hand, the $BS$ correlator receives its main contributions from strange baryons, which are almost equally split between measured and non-measured, given the fact that $\Lambda$ is a charge-neutral particle undetectable by the experiment.

\begin{figure}
	\centering
	\includegraphics[width=\textwidth]{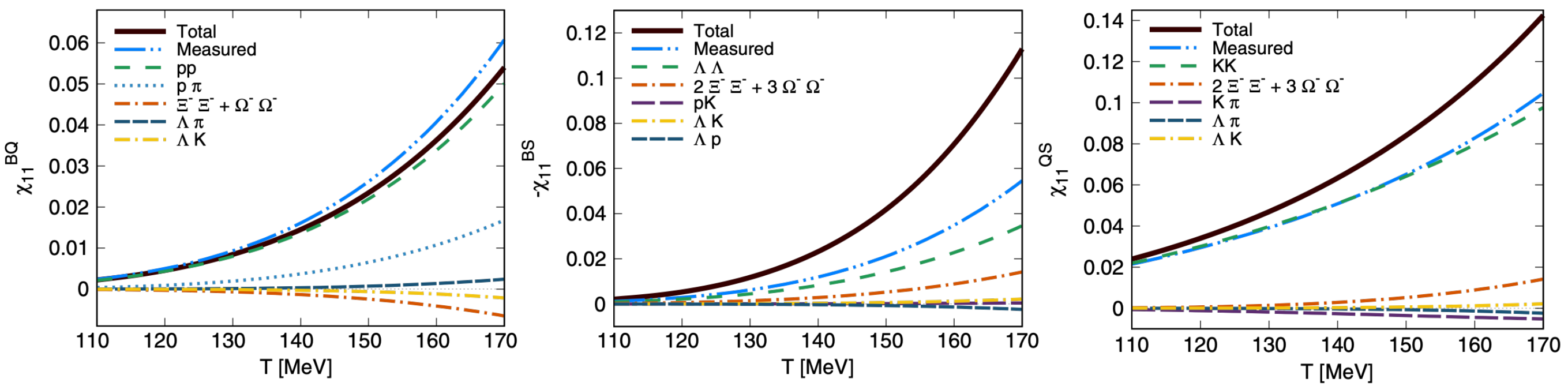}
	\caption{Breakdown of hadronic species contributing to the measurable portion of the cross-correlator as compared to the total value of the correlator \cite{Bellwied:2019pxh}.}
	\label{fig:correl-meas}
\end{figure}

The decomposition in Eq. \eqref{eq:conserved_ch} allows the unique determination of different particle-particle contributions to any cross or diagonal correlator. 
In Fig. \ref{fig:correl-meas}, the breakdown of the measured portion of the single final state hadronic correlations is shown for the cross correlators. 
From this, we see that only a handful of particle-particle correlations contribute to the measured portion of a corresponding observable. 
We, thus, confirm that the $BQ$ and $QS$ correlators are dominated by the contribution from net-proton and net-kaon self-correlations, respectively.
It is also interesting that all correlations between different species yield a very modest contribution, with the only exception being the proton-pion correlator in $\chi_{11}^{BQ}$.
While for the other cross correlators, there is only a small amount of the conserved charge correlations carried by the proton-kaon, kaon-pion, Lambda-pion, and Lambda-kaon correlators in $\chi_{11}^{QS}$, as well as the proton-kaon,  Lambda-kaon and Lambda-proton correlators in $\chi_{11}^{BS}$.

This can be understood when considering that, in this framework, correlations between different particle species can only arise via decay processes.  
When considering the full decay chain from heavier resonances into stable species, if a resonance $R$ has some probability of decaying into stable species $A$ and $B$, then a correlation arises between $A$ and $B$. 
It can be seen from Eq. \eqref{eq:BQ_corr}, in order to have a nonzero correlation, both probabilities must be nonzero. 
For the same reason, correlations between different baryons arise, despite the fact that there exists no decay mode with more than one (anti-)baryon.
Finally, since both $\Xi^-$ and $\Omega^-$ carry all three conserved charges, they contribute to all three correlators through their self-correlations, and their contribution is not necessarily negligible.

In theory and experiment, it is customary to utilize ratios of fluctuations to cancel the volume dependence to leading order.
For this reason, we focused on the ratio $\chi_{11}^{BS}/\chi_2^S$ to capture the baryon-strangeness proxy from the fluctuations of measured hadrons.
For the $\chi_{11}^{BS}$ correlator, it was often thought that since kaons and protons make up the bulk of particle production, their correlator $\sigma_{pK}$ would be a good proxy.
However, as shown in Fig. \ref{fig:correl-meas}, in reality, the proton-kaon correlation is a poor proxy for the $\chi_{11}^{BS}$ correlator as it gives a negligible contribution.
Instead, what we see from Fig. \ref{fig:correl-meas} is that the variance $\sigma_\Lambda^2$ represents the most sizable contribution to this correlator. 

\begin{figure}
\center
\includegraphics[width=0.49\linewidth]{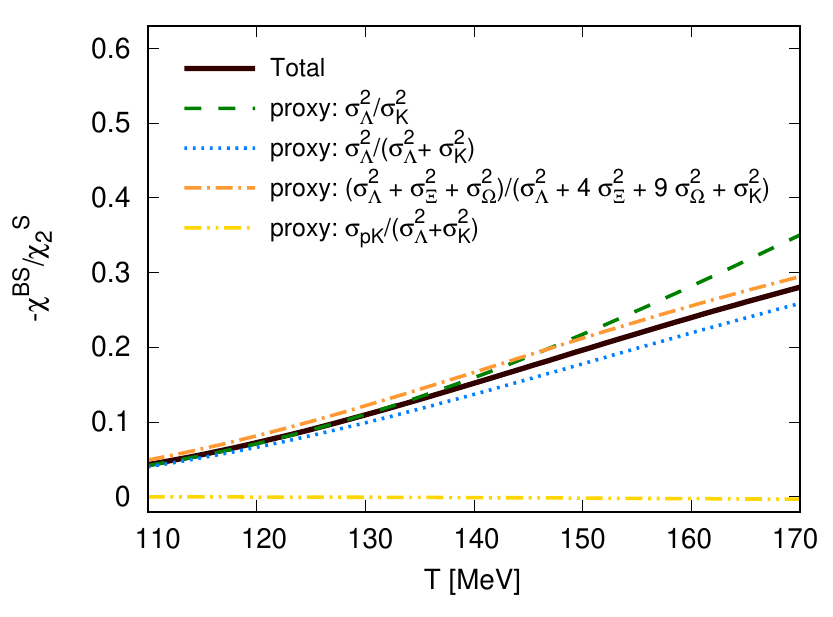}
\caption{Experimentally accessible proxy for cross-correlator of conserved charges $\chi^{BS}_{11}/\chi^S_2$.} 
\label{fig:proxy-BS-QS}
\end{figure}

In Fig. \ref{fig:proxy-BS-QS} the results for the ratio $\chi_{11}^{BS}/\chi_2^S$ are shown.
From this we see that the proposed proxy, $\sigma_\Lambda^2/\sigma_K^2$, does not capture the full result at higher temperatures.
Particularly, the separation increases in the QCD transition region and chemical freeze-out temperatures, which is the area of interest.
A second proxy, $\sigma_\Lambda^2/( \sigma_K^2 + \sigma_\Lambda^2)$ is a much better choice, as it is very close to the full result at all temperatures, including in the vicinity of the QCD transition, as shown in Fig. \ref{fig:proxy-BS-QS}.
In order to provide a full picture, we proposed an additional proxy which includes the contributions from multi-strange hadrons, both in the numerator and denominator: $(\sigma_\Lambda^2 + 2 \sigma_\Xi^2 + 3 \sigma_\Omega^2)/(\sigma_\Lambda^2 + 4 \sigma_\Xi^2 + 9 \sigma_\Omega^2 + \sigma_K^2)$.
This also reproduces the behavior of the full ratio, but since it gives a slight overprediction, it does not improve the situation over the previous one. 
As a final check, we built a proxy from the proton-kaon correlator, $\sigma_{11}^{pK}/( \sigma_K^2 + \sigma_\Lambda^2)$, as shown in Fig. \ref{fig:proxy-BS-QS}. 
We, thus, have shown that this combination is not able to serve as a good proxy.

\begin{figure}
\center
\includegraphics[width=0.49\linewidth]{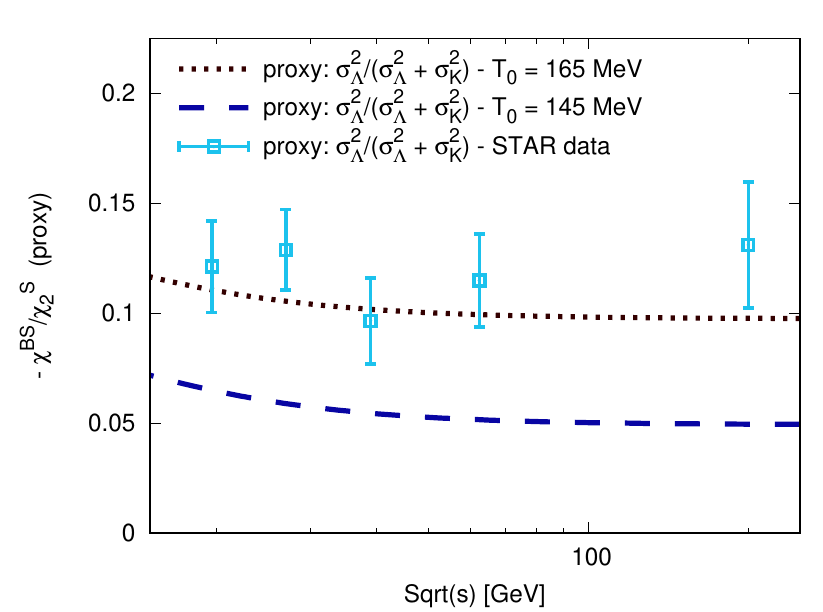}
\caption{Baryon-strangeness correlator from the HRG model, as compared to experimental data from STAR by constructing the proxy as proposed in this study, including the effect of experimental cuts \cite{Bellwied:2019pxh, STAR:2020ddh}.}
\label{fig:proxy-BS-SS-w-cuts}
\end{figure}

Next, since experimental measurements are subject to a finite range of the kinematics, we included the corresponding cuts on a particle basis in order to test how the proxy we constructed compares to the experimental results. 
The versatility of the HRG model allows the incorporation of acceptance cuts in rapidity, $y$, and transverse momentum, $p_T$, that were used in the experiment, as given by the limits of integration in momentum space.
In Fig. \ref{fig:proxy-BS-SS-w-cuts}, the behavior of our proxy along parametrized chemical freeze-out lines is shown as a function of the collision energy $\sqrt{s_\text{NN}}$, which is analogous to the behavior as a function of $\mu_B$.
These chemical freeze-out curves are shifted in $T$ from the parametrization in Ref. \cite{Cleymans:2004pp} with $T_0 = 145, 165$ MeV, in order to capture the entire broad crossover region of QCD.
Overall, we see that the proxy is in good agreement with available experimental data along the freeze-out line with a temperature $T_0 = 165$ MeV.

\section{Constraining Interactions in the Hadron Resonance Gas 
Model via Susceptibilities from Lattice QCD}
Beyond the ideal HRG model as laid out in Sec. \ref{sec:intro}, it is possible to consider an interacting thermal model.
By comparing this model to data from lattice QCD, we can place constraints on interactions in the hadronic phase.
In particular, we utilize the excluded volume HRG (EV-HRG) model, which prevents the overlap of baryons by including a repulsive interaction between those species \cite{Vovchenko:2016rkn,Vovchenko:2017xad,Karthein:2021cmb}.
On the other hand, present already at the level of the ideal HRG model is the attractive interaction between hadrons and resonances that is based on the hadronic spectrum included as input to the model, i.e.  the sum shown in Eq. \eqref{eq:HRG_press}.
By choosing specific combinations of susceptibilities, we can constrain both of these interactions separately.

In order to discuss the EV-HRG model, we will write the pressure in the Boltzmann approximation, such that the partial pressures for particle species $i$ can be written in the following form:
\begin{equation} \label{eq:press_phi}
\centering
    P_i (T, \mu_B, \mu_Q, \mu_S) = d_i \tilde{\phi}(T,m_i) \lambda_i(T,\mu_i), \quad
    \tilde{\phi}(T,m_i) =  \frac{m_i^2 T^2}{2\pi^2} \, K_2(m_i/T).
\end{equation}
The full pressure is then given as the sum of the partial pressures of all hadronic sectors $i$ with various $BQS$ quantum numbers:
\begin{equation} \label{eq:Pfug}
    \centering
    P (T, \mu_B, \mu_Q, \mu_S) = \tilde{\phi_0}(T) + \sum_{i \neq 0} 2 \, \tilde{\phi_i}(T) \cosh \left(\mu_i/T \right),
\end{equation}
where $\mu_i=B_i \mu_B + Q_i \mu_Q + S_i \mu_S$ is the chemical potential of the corresponding $i^{th}$ sector and $\tilde{\phi_0}(T)$ is the $i=0$ term, for charge neutral species. 
The pressure is then separated into non-interacting mesons and interacting baryons and antibaryons: $P = P_M^{\rm id} + P_B^{\rm ev} + P_{\bar{B}}^{\rm ev}$.
\begin{equation} \label{eq:pBev}
\centering
P_{M}^{\rm id}  = \tilde{\phi}_0(T) ~ + \sum_{i \neq 0, \, i \in M} 2 \, \tilde{\phi}_i(T) \, \cosh(\mu_i/T), \quad
P_{B(\bar{B})}^{\rm ev}  =  \sum_{i \in B} \tilde{\phi}_i(T) \, \exp(\pm \mu_i/T) \,
\exp\left( \frac{- b \, P_{B(\bar{B})}^{\rm ev}}{T} \right)
\end{equation}
Here, $i \in M$ corresponds to mesons~($B_i = 0$), $i \in B$ corresponds to baryons~($B_i = 1$), $b$ is the baryon excluded volume parameter, and $\tilde{\phi}(T)$ is given in Eq. (\ref{eq:press_phi}).
Equation~\eqref{eq:pBev} can be solved in terms of the Lambert W function:
\begin{equation}\label{eq:pBevW}
P_{B(\bar{B})}^{\text ev}  = \frac{T}{b} \, W[\varkappa_{B(\bar{B})}(T,\mu_B,\mu_Q,\mu_S)], 
\varkappa_{B(\bar{B})}(T,\mu_B,\mu_Q,\mu_S) = b \, \sum_{i \in B} \tilde{\phi}_i(T) \, \exp(\pm \mu_i/T).
\end{equation}

The hadronic spectrum is studied via the inclusion of different numbers of resonant states in the model.
The most straightforward choice is to utilize the experimentally established states, as measured by the Particle Data Group (PDG).
However, it has been shown that these states may not be enough to describe the full hadronic spectrum \cite{Alba:2017mqu}.
The PDG classifies particles based on how well established those hadronic states are based on an assignment of a number of stars (*). 
The **** states are those which are unambiguously known, such as protons or $\Delta(1232)$ resonances. 
Conversely, the * states are the least established, for example the $\Delta$(1750) or other high-mass resonances. 
The three hadronic lists under study here are as follows: 
PDG2016 -- ordinary hadronic list (*** $-$ ****) from the 2016 Particle Data Booklet;
PDG2016+ -- hadronic list with additional states beyond those well-established ~(*$-$****);
Quark Model~(QM) -- all states predicted by the Quark Model.
The lists including further hadronic states were introduced and described in detail in \cite{Alba:2017mqu, Alba:2020jir}.
It is also useful to note that the QM list used here was updated in Ref. \cite{Karthein:2021cmb}. 

Finally, with this model setup, we can identify the combinations of susceptibilities that will allow us to constrain the hadronic spectrum and excluded volume, $b$, separately.
We start with second order susceptibility ratios that will be independent of the excluded volume parameter, $b$, and therefore, is only sensitive to the partial pressures and the hadron spectrum.
By calculating them explicitly one obtains
\begin{align} \label{eq:chi11BQ}
\frac{\chi_{11}^{BQ}}{\chi_2^B} & = \frac{\sum_{j \in \rm{sectors}} \, B_j \, Q_j \, \tilde{\phi}_j (T) }{\sum_{j \in \rm{sectors}} \, B_j^2 \, \tilde{\phi}_j (T)}, \\
 \label{eq:chi11BS}
\frac{\chi_{11}^{BS}}{\chi_2^B} & = \frac{\sum_{j \in \rm{sectors}} \, B_j \, S_j \, \tilde{\phi}_j (T) }{\sum_{j \in \rm{sectors}} \, B_j^2 \, \tilde{\phi}_j (T)}.
\end{align}

On the other hand, we can place constraints on the excluded volume via fourth-to-second order susceptibility ratios.
The following three ratios are all equal in this EV-HRG model and are sensitive to the EV parameter $b$:
\begin{align} 
\frac{\chi^B_4}{\chi^B_2} = \frac{\chi^{BS}_{31}}{\chi^{BS}_{11}} =  \frac{\chi^{BQ}_{31}}{\chi^{BQ}_{11}} 
& = \frac{1 - 8 \, W(\varkappa_B) + 6 [W(\varkappa_B)]^2}{[1 + W(\varkappa_B)]^4} \nonumber \\
& = 1 - 12 \varkappa_B + O(\varkappa_B^2).
\end{align}    
In the absence of EV repulsive interactions, $\varkappa_B = 0$, and $\frac{\chi^B_4}{\chi^B_2} = \frac{\chi^{BS}_{31}}{\chi^{BS}_{11}} = \frac{\chi^{BQ}_{31}}{\chi^{BQ}_{11}} = 1$, without any effect from the presence of additional hadronic states. 
Thus, the suppression of these ratios relative to unity can be used to estimate a valid range of the EV parameter $b$.
Furthermore, the equality of these three ratios within the model gives us an idea about the limits of validity of the model, as shown by the point at which the ratios are no longer equal in the lattice QCD data.

\begin{figure*}
    \centering
    \includegraphics[width=0.44\linewidth]{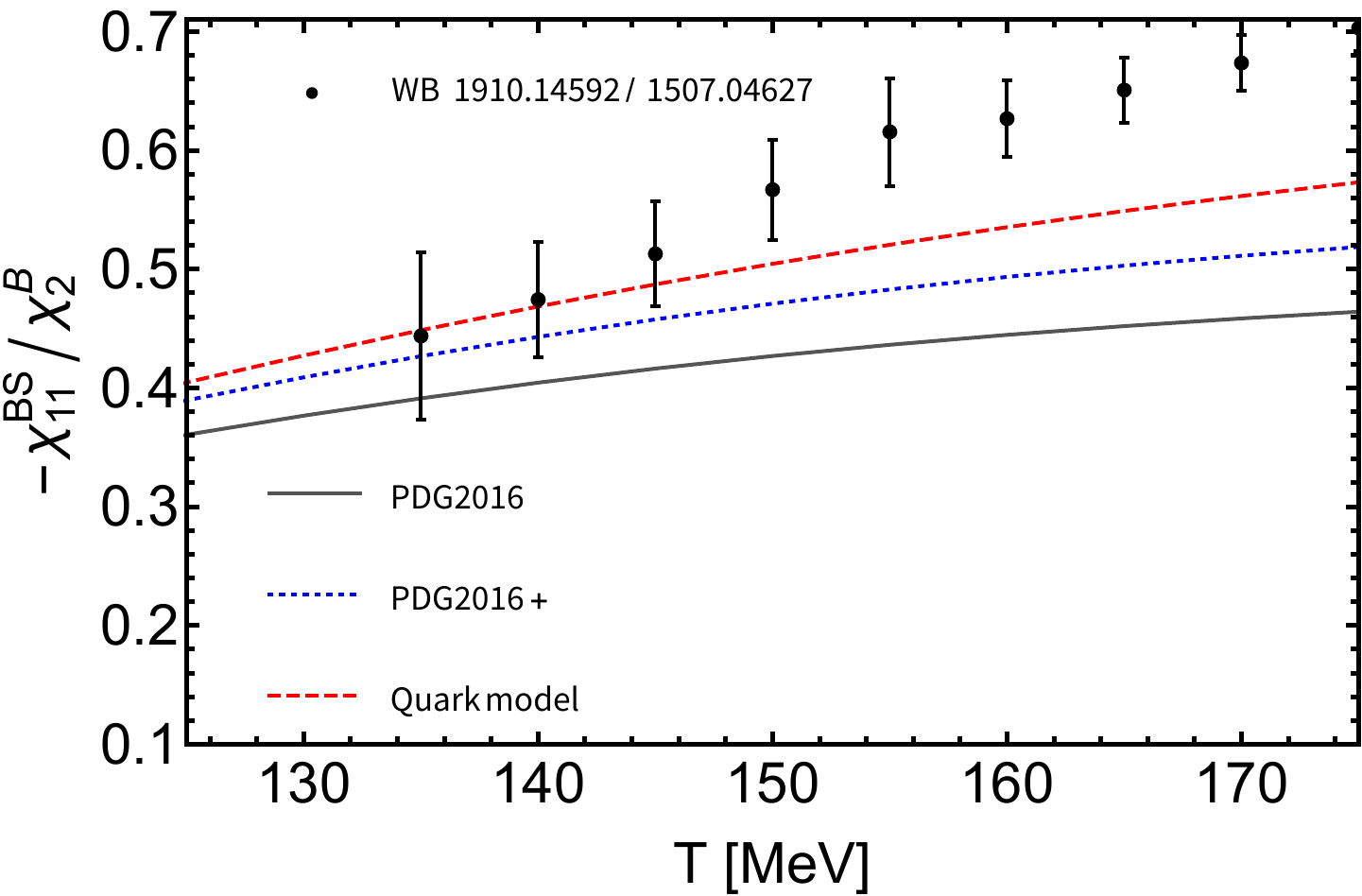}
    \includegraphics[width=0.44\linewidth]{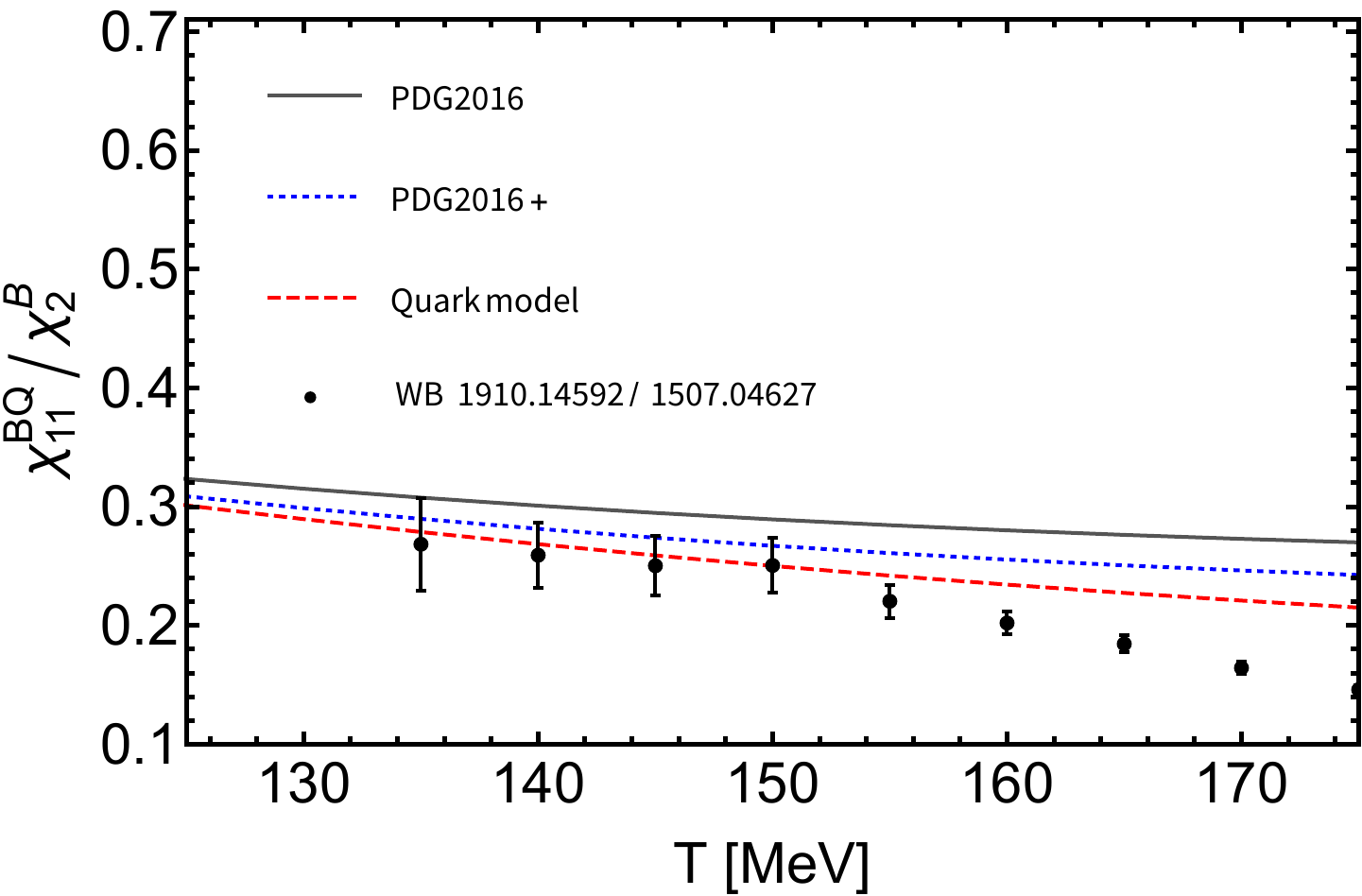}
    \includegraphics[width=0.45\linewidth]{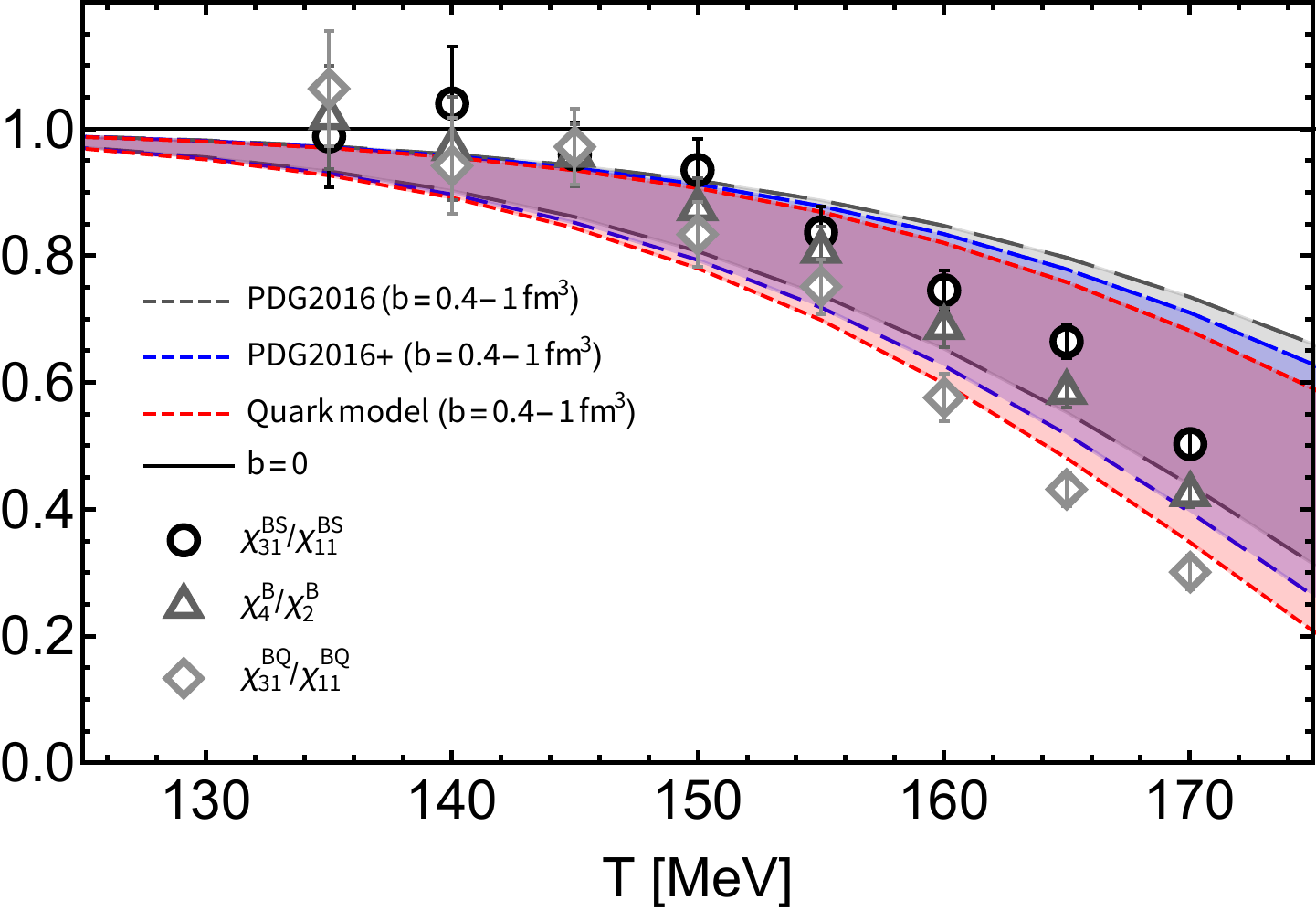}
    \caption{Temperature dependence of susceptibility ratios in the EV-HRG model (colored lines) and from lattice QCD (grayscale points) \cite{Bellwied:2015lba,Bellwied:2019pxh}. Top: second order susceptibility ratios from lattice QCD \cite{Borsanyi:2018grb} and EV-HRG model \cite{Karthein:2021cmb} insensitive to the excluded volume parameter $b$. Bottom: fourth-to-second order susceptibility ratios, insensitive to the hadron spectrum. Here the colored bands correspond to a range of $b: 0.4-1$ fm.}
\label{EV-HRG}
\end{figure*}

Figure \ref{EV-HRG} shows the results within the EV-HRG for these special combinations of susceptibilities.
In the top panel, one can see that the number of states included in the PDG2016 list is not enough to reproduce the results from lattice QCD.
More states are require to sufficiently describe the data either with the PDG2016+ or Quark Model lists for both strange and charged baryons.
In the bottom panel of Fig. \ref{EV-HRG}, we see that we can constrain the EV parameter, $b$, with the three fourth-to-second order susceptibilities.
From the lattice QCD data and the EV-HRG curves, we see that the qualitative behavior of the three ratios is similar over this temperature range.
Looking closer, we see, quantitatively, that for $T \leq 150$ MeV the three ratios are equal. 
Significant differences in the susceptibility ratios appear in the lattice data for $T \gtrsim$~160~MeV.
This may suggest a breakdown of the model in the transition region where the system becomes deconfined.
On the other hand, we suggest that these differences may also reflect a flavor-dependence of baryon excluded volumes.
In our EV-HRG model, all baryons have EV parameter $b$ by construction, leading to the equality of these ratios, i.e. overlap in the curves.
However, if for example strange baryons were to carry a different excluded volume, then $\chi_{31}^{BS}/\chi_{11}^{BS}$ should show smaller deviations from unity than the other ratios.
This indicates that the strange baryons, as represented here in Fig. \ref{EV-HRG} may in fact have a smaller volume than their non-strange counterparts.
One can understand this based on the fact that interactions between strange baryons are mediated by the exchange of heavier mesons like $\phi$, while nucleon-nucleon interactions proceed via the exchange of lighter mesons like $\sigma$.

\section{Conclusions}
In this manuscript, we explored the study of fluctuations and correlations in the system of strongly-interacting matter by utilizing available experimental measurements and lattice QCD data.
The leading contributions to correlators of conserved charges were determined by performing an analysis within the Hadron Resonance Gas (HRG) model along with first-principles lattice QCD calculations.
A comparison of the total amount of a given correlator was compared to experimentally measurable contributions in order to determine the most crucial measurements.
From this analysis, we have constructed a good proxy for a baryon-strangeness correlator via a ratio of second order susceptibilities.
The proxy makes use of the smallest number of hadronic correlations, namely the variances $\sigma^2_K$ and $\sigma^2_\Lambda$.
We have shown that simply a combination of these two provided a sufficient description of a baryon-strangeness cross correlation.
It is remarkable to note that the addition of multi-strange baryons to this proxy does not improve the existing agreement.
On the other hand, we considered extensions to the HRG model and proposed ratios of susceptibilities to constrain the interactions in the model.
This particular EV-HRG model provide a minimalistic extension of the ideal HRG model by only allow interactions of baryons and anti-baryons with themselves, since the other combinations are not well motivated phenomenologically.
From combinations of second order susceptibilities, we have seen that the most established states from the PDG are not enough to encapsulate the full hadronic spectrum, as compared to lattice QCD results.
Additionally, we have placed constraints on the EV parameter via fourth-to-second order ratios that are largely insensitive to the hadronic spectrum.
In this case, we see that a volume of $b \sim 0.4 - 1 \mathrm{fm}^3$ is allowed by the lattice data.
Furthermore, the HRG model exhibits its breakdown point, as shown in Fig. \ref{EV-HRG}, where the lattice data no longer agrees, as suggested in the model approach.

\ack{The author would like to thank the CSQCD IX organizers for the invitation and invigorating conference experience. She also gratefully acknowledges support from the National Science Foundation under Award No. 2138063.}

\section*{References}
\bibliography{all}

\end{document}